\documentclass[prl,twocolumn,superscriptaddress,showpacs]{revtex4}
\usepackage{amsmath}
\usepackage{amsfonts}
\usepackage{amssymb}
\usepackage{graphicx}

\begin{document}

\title{A simple nearest-neighbor two-body Hamiltonian system for which \\
the ground state is a universal resource for quantum computation}
\author{Stephen D. Bartlett}
\affiliation{School of Physics, The University of Sydney, Sydney,
New South Wales 2006, Australia}%
\author{Terry Rudolph}
\affiliation{Department of Physics, Imperial College
London, London SW7 2BW, United Kingdom}%
\affiliation{Institute for Mathematical Sciences, Imperial College
London, London SW7 2BW, United Kingdom}%

\date{24 October 2006}

\begin{abstract}

We present a simple quantum many-body system -- a two-dimensional
lattice of qubits with a Hamiltonian composed of nearest-neighbor
two-body interactions -- such that the ground state is a universal
resource for quantum computation using single-qubit measurements.
This ground state approximates a cluster state that is
\emph{encoded} into a larger number of physical qubits. The
Hamiltonian we use is motivated by the projected entangled pair
states, which provide a transparent mechanism to produce such
approximate encoded cluster states on square or other lattice
structures (as well as a variety of other quantum states) as the
ground state.  We show that the error in this approximation takes
the form of independent errors on bonds occurring with a fixed
probability.  The energy gap of such a system, which in part
determines its usefulness for quantum computation, is shown to be
independent of the size of the lattice. In addition, we show that
the scaling of this energy gap in terms of the coupling constants of
the Hamiltonian is directly determined by the lattice geometry.  As
a result, the approximate encoded cluster state obtained on a
hexagonal lattice (a resource that is also universal for quantum
computation) can be shown to have a larger energy gap than one on a
square lattice with an equivalent Hamiltonian.

\end{abstract}


\maketitle


The state of a quantum many-body system can serve as a universal
resource for quantum computing, where computation proceeds through
single-qubit measurements alone~\cite{Rau01,Rau03,Nie05}. This
observation raises the intriguing possibility that there exist
physical systems which are ``naturally'' quantum computers.  More
precisely, one may ask whether there exist quantum many-body systems
which, when cooled sufficiently close to the ground state, can be
used for quantum computation by simply making individual
measurements on the constituent particles.

It is straightforward to write down a Hamiltonian for a many-body
spin system for which this is the case -- one needs only to take the
(negative) sum of the stabilizer operators corresponding to, say, a
square lattice cluster state~\cite{Rau05}.  This Hamiltonian is
somewhat unsatisfactory, however, because it involves five-body
interactions, and fundamental interactions are strictly between two
bodies.  Using the results of Haselgrove \emph{et al.}~\cite{Has03},
Nielsen proved the following negative result: a cluster state
suitable for quantum computation cannot arise as the exact ground
state of any Hamiltonian involving only local two-body
interactions~\cite{Nie05}. However, Oliveira and
Terhal~\cite{Oli05}, building on the results of Kempe, Kitaev and
Regev~\cite{Kem04}, demonstrated that cluster states (and other such
states that are universal) can be \emph{approximated} by the ground
state of a local two-body Hamiltonian. The key idea of their result
is to make use of ``mediating'' ancilla qubits to create an
effective many-body coupling out of two-body interactions.

Here, we provide a simple and explicit scheme for cluster-state
quantum computation using the ground state of a Hamiltonian
consisting of only two-body interactions.  Our approach, which
provides an alternative to the use of mediating systems as
in~\cite{Kem04,Oli05}, is motivated by the \emph{projected entangled
pair states} (PEPS) as proposed by Verstraete and
Cirac~\cite{Ver04}. Using PEPS, it was demonstrated that the cluster
state can be obtained through a projection of a number of virtual
qubits prepared in maximally-entangled states down to a lesser
number of physical qubits.  We turn this construction around, and
consider how, through the use of nearest-neighbor two-body
Hamiltonians, we may effect a similar projection -- but in this
instance of a number of physical qubits down to a lesser number of
\emph{logical} qubits. The system we investigate is gapped, as the
energy difference between the ground and first-excited states is
independent of the size of the lattice.  The low-energy theory is
described by an effective five-body interaction Hamiltonian that is
precisely equal to the (negative) sum of the stabilizer operators
acting on the logical qubits. That is, a cluster state encoded as a
logical state is approximated by the ground state of a larger number
of physical qubits.  This encoding provides a very transparent
mechanism for achieving the effective many-body coupling out of
two-body interactions, which can be of Ising and Heisenberg form,
and is naturally extendible to other PEPS constructions. Crucially,
despite using logical qubits consisting of several physical qubits,
measurement-based quantum computation can proceed on the encoded
cluster state using only \emph{single}-qubit measurements.

We also investigate the usefulness of this ground state for
measurement-based quantum computation. Specifically, we analyze the
errors associated with the fact that the ground state only
approximates the encoded cluster state, and demonstrate that the
(fixed) energy gap is directly determined by the lattice geometry.

\begin{figure}
\begin{centering}
\includegraphics[width=70mm]{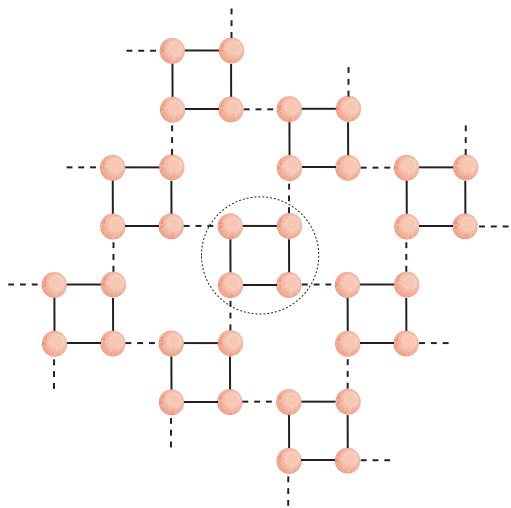}
\end{centering}
\caption{Schematic of the spin lattice, with couplings.  Solid lines
represent an Ising-type interaction, as in the Hamiltonian of
Eq.~\eqref{eq:Ising}.  The dashed lines represent a coupling as in
the Hamiltonian of Eq.~\eqref{eq:Bonds}.  Groups of four physical
qubits bound by the Ising interaction, as within the circle, form a
site -- a logical qubit.} \label{fig:clusterlattice}
\end{figure}

The PEPS on a square lattice~\cite{Ver04} consists of a
2-dimensional lattice of \emph{virtual} qubits following the
Archimedian tiling $4.8.8$ (also known as a ``CaVO''
lattice)~\cite{Ric04}, as shown in Fig.~\ref{fig:clusterlattice}.
Quadruples of virtual qubits on the vertices of the 4-vertex tiles
(such as those circled) form \emph{sites}.  Pairs of virtual qubits
at neighboring sites, connected by dashed-line bonds in
Fig.~\ref{fig:clusterlattice}, are prepared in the two-qubit cluster
state~\footnote{Using PEPS, it is standard to make use of the
singlet state rather that the two-qubit cluster state. As both
states are maximally-entangled and therefore related by local
unitaries, these approaches are equivalent.}
\begin{equation}\label{eq:C2}
    |C_2\rangle = \frac{1}{\sqrt{2}}\bigl(|0\rangle|{+}\rangle +
    |1\rangle|{-}\rangle \bigr)\,.
\end{equation}
The four virtual qubits at each site are projected down to a single
\emph{physical} qubit using the projector
\begin{equation}\label{eq:PEPSproj}
    P = |0\rangle\langle 0000| + |1\rangle\langle 1111| \,.
\end{equation}
The resulting state of the physical qubits on the lattice is the
cluster state~\cite{Ver04}.

In our approach, we choose a nearest-neighbor two-body Ising-type
Hamiltonian such as to effectively implement the above projection on
four physical, rather than virtual, qubits.  Consider a lattice of
physical qubits with the same lattice structure as the virtual
qubits above, as in Fig.~\ref{fig:clusterlattice}.  We label sites
of four physical qubits by a Greek character, e.g., $\mu$, and
define $S$ to be the set of all sites.  Qubits at each site interact
with nearest neighbors via a site Hamiltonian $H_S$ which is of
Ising form
\begin{equation}
  \label{eq:Ising}
  H_S = - \sum_{\mu \in S} \sum_{i \sim i'}
  \sigma^z_{(\mu,i)}\otimes \sigma^z_{(\mu,i')}\,,
\end{equation}
where the first sum is over all sites, $i\sim i'$ denotes pairs of
neighboring qubits at site $\mu \in S$ connected by solid-line bonds
in Fig.~\ref{fig:clusterlattice}, and $\sigma^z_{(\mu,i)}$ is the
Pauli $z$-operator for the physical qubit $i$ at site $\mu$.

The bonds between sites in Fig.~\ref{fig:clusterlattice} indicate a
different two-body interaction, given by a two-body Hamiltonian of
the form
\begin{equation}
  \label{eq:Bonds}
  V=-\!\!\!\sum_{(\mu,i)\sim (\nu,j)}
  \left(\sigma^z_{(\mu,i)}\otimes\sigma^x_{(\nu,j)}
  +\sigma^x_{(\mu,i)}\otimes\sigma^z_{(\nu,j)}\right)\,,
\end{equation}
where $\sigma^x$ is a physical Pauli $x$-operator, and $(\mu,i)\sim
(\nu,j)$ denotes pairs of neighboring qubits connected by
dashed-line bonds in Fig.~\ref{fig:clusterlattice}.

The total Hamiltonian for the lattice is given by the sum of these
terms, $H = g H_S + \lambda V$, where $g$ and $\lambda$ have units
of energy.  To investigate the spectrum of this Hamiltonian, we will
use perturbation theory in $\lambda/g$, with $gH_S$ the unperturbed
Hamiltonian.

We first investigate the spectrum of the unperturbed Hamiltonian
$gH_S$.  At a single site, there are three energy levels.  The
ground-state is degenerate, two-dimensional, and spanned by the
states
\begin{equation}
  |0000\rangle \equiv |\mathbf{0}_L\rangle\,, \quad
  |1111\rangle \equiv |\mathbf{1}_L\rangle\,.
\end{equation}
The ground state space of the unperturbed Hamiltonian, then, can be
viewed as a \emph{logical qubit}.  The energy of this ground state
space is $-4g$. The first excited state is 12-fold degenerate, and
has a energy of $0$. The second excited state is 2-fold degenerate
and has a energy of $4g$.

Consider a lattice as in Fig.~\ref{fig:clusterlattice} consisting of
$N_S$ sites with periodic boundary conditions.  The spectrum of the
unperturbed Hamiltonian $gH_S$ is straightforward.  The ground-state
space has energy $E_0^{(0)} = -4gN_S$, is $2^{N_S}$-dimensional, and
is spanned by all logical states of $N_S$ qubits.  We denote this
space $\mathcal{H}_L$. A convenient basis for this ground state
space is as follows: let $|C\rangle$ denote the \emph{logical
cluster state}, and let $|C\{\alpha,\beta,\ldots\}\rangle$ denote
the logical cluster state with logical $Z$-errors on the sites
$\alpha,\beta,\ldots$.  (The logical $Z$ operator is $Z \equiv
|\mathbf{0}_L\rangle\langle \mathbf{0}_L| -
|\mathbf{1}_L\rangle\langle \mathbf{1}_L|$.)  The set of states $\{
|C\rangle,|C\{\alpha\}\rangle,|C\{\alpha,\beta\}\rangle,\ldots \}$,
running over logical $Z$-errors at all possible sites, forms a basis
for this ground state space.

The first-excited space is $(12N_S\cdot 2^{N_S-1})$-dimensional, and
has energy $E_1^{(0)} = -4g(N_S-1)$.  Thus, for the unperturbed
Hamiltonian $gH_S$, the gap from the ground to first-excited space
is $4g$.  The second-excited space has energy $E_2^{(0)} =
-4g(N_S-2)$. These energies will serve as the zeroth-order energies
in perturbation theory for the total Hamiltonian.


We now turn to perturbation theory.  The term $\lambda V$ in the
Hamiltonian, representing the coupling on the bonds, breaks the
degeneracy of the ground state space; however, as we will see, it
only does so at a fourth-order perturbation in $\lambda/g$.  Let
$\Pi_L$ be the projection onto the ground state space of the
unperturbed Hamiltonian $gH_S$, i.e., onto the ``logical'' space. It
is straightforward to show that $\Pi_L V \Pi_L = 0$, and thus there
is no first-order correction to the energies.  It is also
straightforward to show that
\begin{equation}
  \Pi_L V^2 \Pi_L = 2 \times (\text{number of bonds}) = 4N_S \Pi_L\,.
\end{equation}
(This constant arises from the fact that each Pauli term in $V$
squares to the identity.)  Thus, there is a constant second-order
correction to the ground-state energy -- a shift -- given by
\begin{equation}
    \lambda^2 E^{(2)}_0 = \frac{4N_S\lambda^2}{E_0^{(0)} - E_1^{(0)}}
    = -N_S\frac{\lambda^2}{g} \,.
\end{equation}
There is no third-order correction to the ground state energy,
following from the fact that $\Pi_L V^3 \Pi_L = 0$.  Finally, at
fourth-order, the degeneracy is broken.  To calculate the correction
to the energies, we will make use of two convenient properties of
our encoding. First, we note that the product of all four $\sigma^x$
operators on a single site (one on each physical qubit) results in a
\emph{logical} $X$ operator
\begin{equation}
    X_\mu = |\mathbf{0}_L\rangle\langle \mathbf{1}_L| +
    |\mathbf{1}_L\rangle\langle \mathbf{0}_L| =
    \sigma^x_{(\mu,1)}\sigma^x_{(\mu,2)}\sigma^x_{(\mu,3)}\sigma^x_{(\mu,4)}\,.
\end{equation}
Second, we note that a \emph{single} $\sigma^z$ operator acting on
\emph{any} of the four qubits at a site is equivalent to a logical
$Z$ at that site.  With these facts, we find that
\begin{equation}
  \label{eq:FourthAsStabilizers}
  \Pi_L V^4 \Pi_L - (\Pi_L V^2 \Pi_L)^2
  = 4! \sum_{\mu \in S} K_\mu\,,
\end{equation}
where
\begin{equation}\label{eq:ClusterStabilizer}
    K_\mu \equiv X_\mu \prod_{\nu\sim\mu}Z_\nu \,,
\end{equation}
is a stabilizer of the logical cluster state.  In this expression,
$X_\mu$ is a logical $X$ acting at the site $\mu$, and $Z_\nu$ is a
logical $Z$ acting on a site that is connected to $\mu$ by a bond.

The operator $\Pi_L V^4 \Pi_L$ is already diagonal in our chosen
basis (the states of the form $|C\{\alpha,\beta,\ldots\}\rangle$).
The fourth-order energy corrections for these states are determined
by their eigenvalues, which are straightforward to calculate using
the properties of stabilizers. First, we note that the cluster state
$|C\rangle$ is an eigenstate of all stabilizers in the sum
Eq.~\eqref{eq:FourthAsStabilizers} with eigenvalue $+1$, and thus
the fourth-order correction for the energy associated with this
state is
\begin{equation}
    \lambda^4 E^{(4)}_{|C\rangle}
    = \frac{4!N_S\lambda^4}{(E_0^{(0)} - E_1^{(0)})^2(E_0^{(0)}-E_2^{(0)})}
    = -N_S\frac{3}{16}\frac{\lambda^4}{g^3} \,.
\end{equation}
Next, consider a state $|C\{\alpha\}\rangle = Z_\alpha|C\rangle$, a
cluster state with a single $Z$-error at the site $\alpha$.  This
state is also an eigenstate of all stabilizers in the sum
Eq.~\eqref{eq:FourthAsStabilizers} with eigenvalue $+1$
\emph{except} the stabilizer $K_\alpha$ which has eigenvalue $-1$.
(This result is due to the fact that $X_\alpha$ and $Z_\alpha$
anticommute.)  Thus, to fourth order, the spectrum of the
Hamiltonian $H = gH_S + \lambda V$ is as follows:  the component of
the ground state in $\mathcal{H}_L$ is the cluster state
$|C\rangle$, with energy
\begin{align}
    E_0 = -4gN_S - N_S \frac{\lambda^2}{g}
    -N_S\frac{3}{16}\frac{\lambda^4}{g^3} \,.
\end{align}
The $n$th excited space is $\binom{N_S}{n}$-dimensional.  The
component of this space in $\mathcal{H}_L$ is spanned by states
obtained from $|C\rangle$ with $n$ logical $Z$ errors.  These states
have energy $E_n = E_0 + n\Delta$, up to $n=N_S$, where
\begin{equation}
  \Delta \equiv E_1-E_0 = \frac{3}{16}\frac{\lambda^4}{g^3} \,.
\end{equation}
(Above $n=N_S$, there is a large gap of order $6g$ to the first
manifold of illogical states, which are those that include a single
physical $\sigma^x$ error.)  We note that the gap $\Delta$ is
independent of the size of the lattice.  Intuitively, then, one may
associate logical $Z$ errors on any site with a fixed energy $\Delta
= 3\lambda^4/(16g^3)$ each.


Although the ground-state energy degeneracy of the unperturbed
Hamiltonian $gH_S$ is not broken until fourth order, the lowest
energy eigenstates are corrected even at first order as they will
include terms from the higher-energy eigenspaces of $gH_S$.  A
simple counting of the states involved, and the magnitude of the
components, will allow us to calculate the overlap of the
first-order-corrected ground state with the exact cluster state.  It
will also allow for a simple determination of an error probability
per bond associated with this correction.

At first order, a state in the ground-state space (we will use
$|C\rangle$ as an example, but this calculation holds for any state)
will be corrected to include components from states $|\psi\rangle$
in the first excited space of $gH_S$ whenever $\langle
\psi|V|C\rangle$ is non-zero. We note that every term of the form
$\sigma^x\otimes \sigma^z$ or $\sigma^z\otimes \sigma^x$ from $V$
acting across a bond will connect $|C\rangle$ to a unique excited
state, all of which form an orthogonal set; thus, summing over all
the terms in $V$, there will be $4N_S$ orthogonal states, denoted
$|k\rangle$, for which $\langle k|V|C\rangle \neq 0$.  The first
order corrected ground state $|E_0\rangle$ is
\begin{equation}\label{eq:FirstState}
    |E_0\rangle = \Bigl( 1 + \frac{N_S}{4}\frac{\lambda^2}{g^2}\Bigr)^{-1/2}
    \Bigl( |C\rangle - \frac{\lambda}{4g}\sum_{k=1}^{4N_S}|k\rangle \Bigr) \,.
\end{equation}

We now consider how measurement-based quantum computation can
proceed using such a state.  First, we need to verify that single
qubit measurements alone suffice to perform universal quantum
computation with an encoded cluster state.  The logical cluster
qubits consist of four physical qubits, and cluster-state
computation requires making projective measurements on the logical
qubits onto entangled superpositions of physical qubits the form
$|\mathbf{0}_L\rangle\pm e^{-i\alpha}|\mathbf{1}_L\rangle$.  By
performing separate projective measurements on 3 of the 4 physical
qubits in the $|\pm\rangle$ basis, the logical qubit is decoded into
the state of the fourth physical qubit with at worst the addition of
a known $Z$-error if an odd number of $|-\rangle$ outcomes are
obtained~\cite{Bro05}\footnote{The results of~\cite{Wal00} prove
that this result generalizes to any encoding.}. Such errors are
easily compensated for by adapting future measurement bases in the
manner standard for cluster state computing.

Next, consider the effect of the perturbative correction to the
ground state of Eq.~\eqref{eq:FirstState} in terms of its usefulness
for quantum computation.  Clearly, replacing the superposition of
Eq.~\eqref{eq:FirstState} with an incoherent mixture of the state
$|C\rangle$ and the states $|k\rangle$ can only decrease the
usefulness of the ground state for quantum computation.  With this
replacement, the resulting density matrix describing the ground
state is a mixture of cluster states $|C\rangle$ with errors
$\sigma^x\otimes \sigma^z$ (and $\sigma^z\otimes \sigma^x$) applied
to all bonds independently with probability $p \simeq
\lambda^2/(4g)^2$.  Note that this probability is independent of the
size of the lattice, and thus these errors can be viewed as
independent errors in the bonds between sites.

These errors take the system outside of the logical Hilbert space,
because $\sigma^x$ on, say, the first physical qubit on a site maps
logical states to states spanned by $|1000\rangle$ and
$|0111\rangle$. However, these errors manifest themselves in the
single-logical-qubit measurements described above as
\emph{effective} Pauli errors in the measurement. Thus, standard
techniques for fault-tolerance in cluster state quantum
computation~\cite{Rau03b,Nie05b,Ali06} can be applied, provided the
error probability $p\simeq\lambda^2/(4g)^2$ is below some
appropriate threshold.


Errors can also arise due to the finite energy gap, $\Delta =
3\lambda^4/(16g^3)$.  As we are using perturbation theory, we
require $\lambda \ll g$ for the ground state to closely approximate
the encoded cluster state.  However, we emphasize that the energy
scale $\lambda$ need only be small relative to $g$, and not in any
absolute sense. The gap $\Delta$ is in turn small relative to
$\lambda$. Consideration of the relative magnitudes of these energy
scales, in conjunction with determining how cold such a lattice can
be maintained, will determine whether this gap is sufficiently large
to allow for quantum computation. Specifically, consider the effects
of using a finite-temperature thermal state of this Hamiltonian.  As
investigated by~\cite{Kay06}, the thermal state will be a mixture of
cluster states with logical $Z$-errors occurring independently at
each site with probability $p = (1+\exp(\Delta/k_BT))^{-1}$.  The
critical temperature above which the cluster state becomes too noisy
to be useful for quantum computation scales as $k_BT_{\rm crit} \sim
\Delta$~\cite{Rau05,Kay06}, which in this case is $k_BT_{\rm crit}
\sim(\lambda/g)^3\lambda$.  Thus, the critical temperature is
determined by the energy scale $\lambda$ \emph{and} the order of
perturbation theory at which the degeneracy is broken. (For the
square lattice, the latter is four, leading to the $(\lambda/g)^3$
dependence.)

In relation to this point, we note that similar techniques to those
presented here can be used to construct systems for any type of
cluster state, not just on a square lattice, and a large number of
other quantum states on a graph. This result follows directly from
the origins of our method in the PEPS formalism.  To generalize the
above method, the number of physical qubits at each site must equal
the number of bonds, i.e., the number of other sites that are
directly connected to that site. For example, line- or ring-clusters
can be created using only two physical qubits per site, a hexagonal
lattice cluster state using three per site, and a cubic lattice
cluster state using six per site.  We note that the number of qubits
per site determines the order in perturbation theory at which the
ground-state degeneracy is broken.  Continuing our argument from the
previous paragraph, it is therefore interesting to consider the use
of a hexagonal lattice cluster state of this form. Such a cluster
state is a universal resource for quantum computation~\cite{Van06},
and as the degeneracy is broken at third order, the energy gap (and
thus the critical temperature) behaves as $\Delta_{\rm hex} \sim
\lambda^3/g^2$. This larger energy gap, as compared with the square
lattice scaling of $\Delta_{\rm sq} \sim \lambda^4/g^3$, may make
this state easier to prepare and maintain.

It is worthwhile to consider whether alternative local two-body
Hamiltonians also yield an encoded cluster state as their ground
state, or yield a non-cluster state that is nonetheless universal
for quantum computation.  For example, using a Heisenberg
antiferromagnetic coupling on bonds and alternating sites between
$ZZ$-type and $XX$-type Ising interactions leads to a similar result
as above, where the ground state approximates an encoded cluster
state. (To obtain this result, the choice of logical basis at each
site must be alternated.)  In the dynamical approach to cluster
state creation, it has been shown that the Heisenberg interaction
can be used instead of the Ising interaction~\cite{Bor05}, and
logical encodings for this purpose have been
investigated~\cite{Wei05}.  A future line of research would be to
investigate if lattices with entirely Heisenberg interactions (or
other ``natural'' interactions on sites and bonds) can yield a
computationally-useful state as the ground state. We note that our
lattice of Fig.~\ref{fig:clusterlattice}, and the hexagonal lattice
with three physical qubits per site, have identical structure to the
``CaVO-type'' lattices and star lattices, respectively, for which
the Heisenberg antiferromagnetic coupling can lead to exotic quantum
states~\cite{Wei97,Ric04,Ric04b}.

\begin{acknowledgments}
We thank Chris Dawson, Andrew Doherty and Michael Varnava for
helpful discussions and comments.  SDB acknowledges support from the
Australian Research Council.  TR acknowledges support from the
Engineering and Physical Sciences Research Council of the United
Kingdom, and a University of Sydney Short-Term Visiting Fellowship.
\end{acknowledgments}

\end{document}